\documentstyle[preprint,prl,aps]{revtex}
\begin{document}
\draft

\preprint{
\font\fortssbx=cmssbx10 scaled \magstep2
\hbox to \hsize{
\hbox{\fortssbx University of Wisconsin - Madison}
\hfill$\vcenter{\hbox{\bf MADPH-95-911}
                \hbox{\bf RAL-TR-95-062}
                \hbox{\bf hep-ph/9510457}
                \hbox{October 1995}}$}
}

\title{\vspace*{.5in}
Double gluon fragmentation to {\boldmath$J/\psi$} pairs at the Tevatron}

\author{V. Barger and S. Fleming}
\address{Department of Physics, University of Wisconsin, Madison, WI 53706,
USA}
\author{R.J.N. Phillips}
\address{Rutherford Appleton Laboratory, Chilton, Didcot, Oxon OX11 0QX, UK}

\maketitle
\begin{abstract}

It has been proposed that the large cross sections for prompt
$\psi$, $\psi'$, and $\chi_c$ production at the Fermilab Tevatron
$p\bar p$ collider can be explained by a dominant color-octet term
in the fragmentation function for a gluon to split into quarkonium.
We show that  this mechanism makes testable predictions for
double-quarkonium $\psi\psi$, $\psi\psi'$, $\psi\chi_c$, $\psi\Upsilon$
and $\psi\chi_b$ production, as well as for $W \psi$ production,
using color-octet matrix elements previously determined from
charmonium production data. The $\psi\psi$ signal would already be
measurable at the Tevatron, while the $\psi\chi_c$ and $W\psi$ signals
would be on the edge of present detectability.

\end{abstract}
\pacs{}

\thispagestyle{empty}

\newpage

\narrowtext

Until recently the standard way to calculate $\psi$ production was
the color-singlet model~\cite{cs}, where the $ \psi $ is treated as
a $ c \bar{c}$  pair in a color-singlet ${}^3S_1$ state
with vanishing relative velocity. Although the color-singlet model has
been sucessful in some applications, it predicts rates
that fall orders of magnitude below the data~\cite{tev} when applied to
prompt $\psi$ production at the Fermilab Tevatron via the hard
scattering subprocess $g+g \to \psi +g$.
A proposal to explain this discrepancy is that the dominant
contribution to charmonium production at high transverse
momentum  $p_T$  comes from hard gluon production followed by
fragmentation of the gluon to a $c \bar{c}$ pair in a color-octet
state~\cite{bf,bbl}. The conversion of the $c \bar{c}$ pair to $\psi$ is
nonperturbative and involves the absorption or emission of two soft gluons.
Because there are new, nonperturbative parameters associated with the
color-octet mechanism~\cite{bf,cl,mlm}, the rate for single quarkonium
production cannot be immediately predicted, but rather the experimental
data can be used to determine these parameters.  In the present paper
we show that the color-octet mechanism now makes testable predictions for
double-quarkonium $\psi \psi$, $\psi \psi'$, $\psi \chi_c$,
$ \psi \Upsilon $, and $\psi \chi_b$ production, as well as for
$W \psi$ production at the Tevatron (CM energy $\sqrt s=1.8$ TeV).

Using the factorization approach developed in~\cite{bbl} the
fragmentation function for a gluon to split into a $Q\bar Q$ quarkonium
state $H$ is
\begin{equation}
D_{g \to H} (z, \mu^2) = \sum_n d_{g \to n}(z, \mu^2)
\langle {\cal O}^H_n \rangle \; ,
\label{ff}
\end{equation}
where $z = (E_H+p_H)/(E_g+p_g)$ is the light-cone fraction, $\mu = 2m_Q$
is the renormalization scale, while  $n$ denotes the angular-momentum
quantum numbers ${}^{2S+1}L_J$ and the color quantum number \underline{1}
or \underline{8}. The nonperturbative matrix elements
$\langle {\cal O}^H_n \rangle$ represent the inclusive
probability for forming the state $H$ from the $Q \bar{Q}$
pair~\cite{bbl}; color-singlet values may be calculated from potential
models but color-octet values are unknown.
The relative importance of the different matrix elements may however
be determined by how they scale with $v$, the typical
relative velocity of the heavy quarks in $H$.
The short distance coefficients  $d_{g \to n}$ can be calculated using
perturbation theory in $\alpha_s$; they depend on $z$, $\mu$ and
the quantum numbers indexed by $n$. Note that the dependence on the
quarkonium state $H$ appears only in the factor
$\langle {\cal O}^H_n \rangle$.

The leading-order perturbative calculation of the color-octet
short-distance coefficient gives~\cite{bf}
\begin{equation}
d_{g \to\underline{8}\, ^3S_1 } = { \pi \alpha_s(4m^2_Q) \over 24
m^3_Q} \delta(1 - z)
\equiv K(g \to \underline{8}\, ^3S_1) \delta(1-z) \; ,
\label{oct_coeff}
\end{equation}
with numerical values
$ K(g \to \underline{8} \, ^3S_1) = 0.01\rm\;GeV^{-3}$ for charmonium and
$ K(g \to \underline{8} \, ^3S_1) = 2 \times 10^{-4}\rm\;GeV^{-3}$ for
bottomonium. Here we put $\alpha_s(4m^2_c)=0.26$ and
$\alpha_s(4m^2_b)=0.17$, with $m_c = 1.48 \; \mbox{GeV}$ and
$m_b = 4.88 \; \mbox{GeV}$; these masses are taken from the
Buchm\"uller-Tye potential model calculations of Ref.\cite{eq}.
The leading-order color-singlet term in the fragmentation function,
which does not have a simple analytic form~\cite{by}, is of order
$\alpha_s^3$ and gives much smaller factors
$ K(g \to \underline{1} \, ^3S_1) = 4.3\times 10^{-6}\rm\;GeV^{-3}\;
(3.5\times 10^{-8}\;GeV^{-3})$ for charmonium (bottomonium) cases.  It
is consequently assumed that the color-octet term in the fragmentation
function dominates over the color-singlet term, even though
$\langle {\cal O}^{\psi}_8 \rangle$ is expected to be smaller
than $\langle {\cal O}^{\psi}_1 \rangle$ by a factor of order $v^4$.

Numerical values of the nonperturbative matrix elements are given in
Table~I for the $S$-wave states $\psi$, $\psi'$ and $\Upsilon$,
and in Table~II for the $P$-wave states $\chi_c$ and $\chi_b$.
The color-octet matrix elements
$\langle {\cal O}^{\psi}_8({}^3S_1)\rangle $,
$\langle {\cal O}^{\psi'}_8({}^3S_1)\rangle $ and
$\langle {\cal O}^{\chi_{cJ}}_8({}^3S_1)\rangle $
are determined empirically from Tevatron data~\cite{bf,cl,mlm},
based on calculations using both color-octet and predicted color-singlet
contributions in lowest order. The bottomonium matrix elements
$\langle {\cal O}^{\Upsilon(1S)}_8({}^3S_1)\rangle $,
$\langle {\cal O}^{\Upsilon(2S)}_8({}^3S_1)\rangle $ and
$\langle {\cal O}^{\chi_{bJ}(1P)}_8({}^3S_1)\rangle $ were estimated
from the charmonium values by scaling arguments\cite{cl}. All our
color-octet values follow Ref.\cite{cl} (but we have corrected a misprint
in the value of $\langle {\cal O}^{\chi_{bJ}(1P)}_8({}^3S_1)\rangle $).
The color-singlet matrix elements, shown here for comparison,
are determined from potential model calculations using
a Buchm\"uller-Tye potential \cite{eq}; they are related to the
radial wave function $R$ and its derivative at the origin by
\begin{equation}
\langle {\cal O}^H_1({}^3S_1)\rangle = (2J+1){N_c\over 2\pi}
           |R(0)|^2, \quad H=\psi,\psi ',\Upsilon,
\label{sing_s}
\end{equation}
\begin{equation}
\langle {\cal O}^H_1({}^3P_J)\rangle = (2J+1){3N_c\over 2\pi}
           |R'(0)|^2, \quad H=\chi_c,\chi_b ,
\label{sing_p}
\end{equation}
where $N_c=3$ is the number of colors.

Given these results, predictions can immediately be made for other
processes involving hard gluon production with the gluon fragmenting
to quarkonium.
Two recent papers have considered prompt $\psi$ and $\Upsilon$
production at the LEP $e^+ e^-$ collider via $Z^0 \to q \bar{q} g$
decays~\cite{cky,cho}. In this paper we address tests of the
octet fragmentation mechanism that can be made at the Tevatron
$p\bar{p}$ collider, based on multiple quarkonium production
or on quarkonium production in association with a $W$-boson.

In our calculations we shall neglect the evolution with $\mu^2$ of the
gluon fragmentation function, which greatly simplifies the calculation.
This introduces some error, but including evolution would not necessarily
be an improvement, since naive Altarelli-Parisi evolution does not
respect the phase space constraint $D_{g \to \psi}(z, \mu^2) = 0$ for
$z < M^2_{\psi}/\mu^2$ ~\cite{bdfm}.

\medskip
\leftline{\bf Double-{\boldmath$\psi$} production}

The $p_T$ spectrum of prompt single $\psi$ production is presumed to
be dominated by the color-octet term in the $g \to \psi $
fragmentation function~\cite{mlm} given in Eq.~(\ref{ff}). Assuming
that $q \bar{q} \, , \, gg \to gg$ are the dominant contributing hard
subprocesses, we can predict double-$\psi$ production, with both gluons
fragmenting to $\psi$, through the ratio
\begin{equation}
{ d \sigma (p \bar{p} \to\psi\psi X) \over d p_T(\psi_1) d
p_T(\psi_2)}
\approx {1\over 2} K(g \to \underline{8}\, ^3S_1)
\langle {\cal O}^\psi_8({}^3S_1)\rangle
{d \sigma ( p \bar{p} \to \psi gX) \over d p_T(\psi)}
\delta(p_T(\psi_1)- p_T(\psi_2)),
\label{dcs}
\end{equation}
in the absence of cuts, where the ${1\over 2}$ is a combinatorial factor.
This relation remains true in the presence of $p_T$ and angle cuts,
provided they are applied equally to $g$ and $\psi$.

We note that the $gq\to gq$ and $g\bar q\to g\bar q$ subprocesses are not
in fact completely negligible and contribute about 20\% of single-$\psi$
production with $p_T(\psi) > 4$ GeV at Tevatron energies; this correction
reduces our prediction in Eq.~(\ref{dcs}) by about 20\% and we take it into
account below.  We note also that the differential cross
section on the right-hand side of Eq.~(\ref{dcs}) refers to all single
prompt-$\psi$ production (including both direct $g\to \psi$ fragmentation
and indirect $g\to(\chi_{cJ} ,\psi')\to\psi$ contributions but excluding
$\psi$'s from the decay of $B$-mesons); however, the second $\psi$ on the
left-hand side is produced directly (excluding $\chi_{cJ}$, $\psi'$ and
$B$-meson decays). To obtain the total prompt double-$\psi$ rate, we must
add similar contributions from $\chi_{cJ}$ and $\psi'$ production
(see Table III) multiplied by the corresponding
$\chi_{cJ} \to \psi \gamma$ and $\psi'\to\psi X$ branching ratios,
that give approximately the same $p_T(\psi)$ distributions and increase
the total rate by a further factor $\approx 2.0$. In practice $\psi$ is
usually detected via $ \psi \to \mu^+ \mu^- $ decay, with
branching fraction $ B(\psi \to \mu^+ \mu^-)= 0.0597(25) $;
for this decay mode (indicated by the notation
$\psi_{\mu \mu}$) we obtain the differential cross section for total
prompt double-$\psi$ production as a function of $p_T(\psi)$ shown in
Figure~1. The curve was generated from $q\bar q, gg\to gg$ subprocesses
using the MRSD0 parton distribution
functions\cite{mrs} with renormalization scale and factorization scale
both chosen equal to the transverse momentum of the fragmenting gluon
$p_T(g)\simeq p_T(\psi)$; it includes the enhancements from indirect
contributions.  A pseudorapidity cut of $| \eta | < 0.6$ was imposed
on the $\psi$'s produced.

Integrating Eq.~(\ref{dcs}) and including the factor 2 enhancement
from indirect $g\to(\chi_{cJ},\psi')\to\psi$ contributions, we obtain
\begin{equation}
{\sigma(p \bar{p} \to \psi_{\mu \mu} \psi_{\mu \mu} X) \over
 \sigma(p \bar{p} \to \psi_{\mu \mu} gX)} \approx
 K(g \to \underline{8}\, ^3S_1)
\langle {\cal O}^\psi_8({}^3S_1)\rangle B(\psi \to \mu^+ \mu^-) =
7.5 \times 10^{-6} \; \;,
\label{csrat}
\end{equation}
where the cross sections are defined
with a minimum $p_T$ requirement on all $\psi$.  This equation remains true
after pseudorapidity cuts, provided the same cuts are applied equally to
$\psi$ and to the recoil gluon jet $ g $ in $ p \bar{p} \to \psi gX$. The
observed cross section for single prompt $\psi$ production
at the Tevatron with $p_T(\psi) > 4 \; \mbox{GeV}$ and $|\eta| < 0.6$ is
$\sigma(p \bar{p} \to \psi_{\mu \mu} gX) \approx 24 \; \mbox{nb}$\cite{tev},
from which we infer that the cross section arising from $gg$ final states
is about 19 nb and hence
$\sigma(p \bar{p} \to \psi_{\mu \mu}  \psi_{\mu \mu} X)
\approx 0.14 \; \mbox{pb}$. Thus Eq.~(\ref{csrat}) predicts that of order
10-20 double-$\psi_{\mu\mu}$ production events should be detectable already
with $p_T(\psi)> 4 \;\mbox{GeV}$ and $|\eta(\psi)|<0.6$, for integrated
luminosity of order 100 $\mbox{pb}^{-1}$ now accumulated by each of the
Tevatron detectors.  We note that experimental factors such as
detector efficiency could reduce this number considerably.
On the other hand, if $\psi\to ee$ decays are also detected (there
should be good efficiency for this in events that are already tagged
by one $\psi\to\mu\mu$), the event rate will be increased by a
factor 3 or 4, depending on whether double-$\psi_{ee}$ final states
are included.

\medskip
\leftline{\bf {\boldmath$\psi + {}$}quarkonium production}

All the dependence on the $\psi$ state in Eq.(\ref{dcs}) appears only
in the matrix element $\langle {\cal O}^{\psi}_8 ({}^3S_1)\rangle$.
It is therefore trivial to generalize this equation to the
case where one gluon fragments to a $\psi$ while the other gluon
fragments to some other quarkonium state. We have tabulated the
results in Table~III. Note that the $\chi_{cJ}$ and $\chi_{bJ}$
detection modes and branching fractions remain unspecified, and
that the $g\to H$ fragmentation picture is applicable only for
$p_T(g) > m_H$.  The other branching fractions used are:
$B(\psi' \to \mu^+ \mu^-) = 0.0077(17)$,
$B(\Upsilon(1S) \to \mu^+ \mu^-) = 0.0248(07)$, and
$B(\Upsilon(2S) \to \mu^+ \mu^-) = 0.0131(21)$.
The $\psi +(\chi_c\to\psi\gamma)$ channels
account for about 33\% of the prompt double-$\psi$ production rate,
as already noted above, so they predict about 5
$\psi_{\mu\mu}\psi_{\mu\mu}\gamma$  events for 100~pb$^{-1}$
luminosity; there is some extra loss of efficiency from the need
to measure the photon and reconstruct the $\chi_c$ invariant mass,
but these signals would appear to be approaching detectability.
Similarly, the $\psi +(\psi'\to\psi\pi\pi,\psi\eta)$ channels
account for about 17\% of the prompt double-$\psi$ production rate,
with a few events at present luminosities, but it may well be
impracticable to reconstruct $\psi'$ in these modes because of
backgrounds. The predicted rates for $\psi_{\mu\mu} \psi'_{\mu\mu}$,
$\psi_{\mu\mu}\Upsilon_{\mu\mu}$ and $\psi_{\mu\mu} \chi_b$ production
are all too small to be observable at present. However, we note that
the indirect process $g\to\chi_{bJ}(1P)\to\Upsilon (1S)$ is predicted to
dominate by a factor 3 over direct $g\to\Upsilon (1S)$ fragmentation,
when the relevant $\chi_{bJ}(1P)\to\Upsilon (1S)\gamma$ branching
fractions are folded in, so the net $\Upsilon (1S)$ rate is 4 times
larger than the third row of Table III would suggest.

\medskip
\leftline{\bf {\boldmath$\psi + {}$}gauge boson production}

Another test of  the dominance of the color-octet
term in the gluon fragmentation function is high-$p_T$
$\psi + \mbox{gauge boson}$ production. Requiring the $\psi$ and gauge
boson to have large relative $p_T$ severely restricts the number of
possible production mechanisms.

High-$p_T$ $\psi + \gamma$ production at the Tevatron, via gluon-fusion
and color-singlet-fragmentation mechanisms, was investigated in
Ref.~\cite{rs}; it was found that gluon-fusion was the dominant production
mechanism for all $p_T$ accessible to experiment.
We have investigated the additional contribution from the subprocess
$q \bar{q} \to \gamma g$, with the gluon fragmenting to  $\psi$ via
the color-octet mechanism. We find that this color-octet contribution
falls well below the gluon-fusion contribution for
$p_T < 20$ GeV, essentially because there are many more gluons than
quarks in the proton at the relevant $x$-values
for Tevatron energies.  Hence high-$p_T$ $\psi + \gamma$
production is not sensitive to the color-octet contribution, at $p_T$
values where most events will be seen.

In contrast, high-$p_T$ production of $\psi + W$ offers much
cleaner tests of the color-octet mechanism. It has no gluon-fusion
contribution, and the dominant direct production subprocess is
$q \bar{q} \to W g$ with the gluon fragmenting to $\psi$ via the
color-octet mechanism.  This mechanism contributes the differential
cross section
\begin{equation}
{d \sigma \over d p_T} (p \bar{p} \to W \psi_{\mu \mu}X) \approx
 K(g \to \underline{8}\, ^3S_1)\langle {\cal O}^\psi_8({}^3S_1)\rangle
 B(\psi \to \mu^+ \mu^-){d \sigma \over d p_T}(p \bar{p} \to WgX) \; ,
\label{dcspsiw}
\end{equation}
where $p_T$ refers on the left to $\psi$ and on the right to $g$, for
sufficiently large $p_T$. Rapidity cuts must be applied equally to
$\psi$ and to the recoil gluon jet. Equation (\ref{dcspsiw}) describes only the
direct $g\to\psi$ contribution; indirect $g\to (\chi_{cJ},\psi')\to\psi$
contributions will enhance the rate by a factor 2 as discussed above.
Integrating Eq.~(\ref{dcspsiw}) for $p_T > 5$ GeV and including this
enhancement factor, we obtain
\begin{eqnarray}
\sigma (p \bar{p} & \to & \psi_{\mu \mu}W_{e\nu}X ; p_T(\psi)>5
\;\mbox{GeV})  \approx \nonumber \\
& & 2 K(g \to \underline{8}\, ^3S_1)
\langle {\cal O}^\psi_8({}^3S_1)\rangle  B(\psi \to \mu^+ \mu^-)
\sigma (p \bar{p} \to W_{e\nu} gX; p_T(g) > 5 \; \mbox{GeV})
\approx 10 \; \mbox{fb} \; ,
\label{psiw}
\end{eqnarray}
where both $\psi\to\mu\mu$ and $W\to e\nu$ branching fractions
are included. The value for
$\sigma (p \bar{p} \to W_{e\nu} g;\; p_T(g) > 5 \; \mbox{GeV})$
on the right-hand-side of Eq.~(\ref{psiw}) cannot be taken directly
from experiment, since $W+$jet production has substantial
contributions from $gq(\bar q)\to Wq'(\bar q')$ subprocesses; it was
calculated at tree level for $\sqrt s=1.8$ TeV (divide the rate by
2(4) if $p_T > 10(15)$ GeV instead).
This cross section implies one event per 100 pb$^{-1}$ and is
therefore on the edge of detectability at the Tevatron
at present; however, including the $\psi\to ee$ and $W\to\mu\nu$
decay modes would increase the cross section by a factor 4, and
future increases in luminosity could make it measurable.

\medskip
\leftline{\bf Summary}

The proposal that prompt charmonium production at the Tevatron is
dominated by a color-octet term in the gluon fragmentation function
\cite{bf,bbl} introduces new parameters that can however be determined
from data~\cite{bf,cl,mlm}.  We have shown that this mechanism can now
be tested through its predictions for double-charmonium and charmonium+$W$
production. It is remarkable that indirect $g\to (\chi_{cJ},\psi')\to\psi$
contributions enhance $\psi$ signals by about a factor 2 and
analogous $g\to\chi_{bJ}\to\Upsilon$ contributions, inferred by scaling
arguments\cite{cl}, enhance $\Upsilon (1S)$ production by a factor 4.
We have found that the predicted rate for $\psi\psi$ production
should already be observable at the Tevatron, with the $p_T$
distribution shown in Fig.1; the rates for
$\psi \psi'$, $\psi \chi_c$, $\psi \Upsilon$, and $\psi \chi_b$
production (summarized in Table III) are smaller but the
$\psi+(\chi_c\to\psi\gamma)$ signal could already be detectable too.
We have also investigated  $\psi$ plus gauge boson production
at high $p_T$; $\psi + \gamma$ production proves to be insensitive
to color-octet contributions, but $\psi + W$ production offers a clean
test of the color-octet mechanism, with a predicted rate on the edge
of present observability at the Tevatron.

\medskip
\leftline{\bf Acknowledgments}
RJNP thanks Paul Dauncey for helpful information and the University of
Wisconsin for hospitality during part of this work.  VB thanks the Institute
for Theoretical Physics at the University of California, Santa Barbara for
hospitality during the completion of this work.
This research was supported in part by the U.S.~Department of Energy under
Grant No.~DE-FG02-95ER40896, in part by the University of Wisconsin Research
Committee with funds granted by the Wisconsin Alumni Research Foundation, and
in part by the National Science Foundation under Grant No.~PHY94-07194.


\vbox{

\begin{table}[h]
\caption{Values of the nonperturbative matrix elements
used to calculate $\psi$, $\psi'$ and $\Upsilon$ production at the
Tevatron.  The color-octet matrix elements for charmonium were determined
empirically from CDF data;  the color-octet matrix elements for bottomonium
were determined by re-scaling the corresponding charmonium matrix
elements [5]. The color-singlet matrix elements, shown here for
comparison, were determined from potential model calculations [7].}
\label{swave}
\smallskip
\begin{tabular}{ccccc}
Matrix Element & $\psi$  & $\psi '$  & $\Upsilon(1S)$  & $\Upsilon(2S)$
\\ \hline
$\langle {\cal O}_1({}^3S_1)\rangle $
& 1.2 ${\rm GeV}^3$ &  0.75 ${\rm GeV}^3$ & 9.3 ${\rm GeV}^3$
& 4.5 ${\rm GeV}^3$
\\
$\langle {\cal O}_8({}^3S_1)\rangle $
& 0.012 ${\rm GeV}^3$ &  0.0073 ${\rm GeV}^3$ & 0.01 ${\rm GeV}^3$
& 0.006 ${\rm GeV}^3$
\end{tabular}
\end{table}

\begin{table}[h]
\caption{Values of the nonperturbative matrix elements
used to calculate $\chi_{c}$ and $\chi_b$  production at the Tevatron;
different J-values differ simply through a factor (2J+1).
The color-octet matrix element for $\chi_c$ was determined empirically
from CDF data; the color-octet matrix element for $\chi_b(1P)$
was found by re-scaling the $\chi_c$ matrix element [5].
The color-singlet matrix elements, shown for comparison, were
determined  from potential model calculations [7].}
\label{pwave}
\smallskip
\begin{tabular}{ccc}
Matrix Element & $\chi_{cJ}$ & $\chi_{bJ}(1P)$
\\ \hline
$\langle {\cal O}_1({}^3P_0)\rangle /(2J+1)$
&  0.11 ${\rm GeV}^5$  & 2.0 ${\rm GeV}^5$
\\
$\langle {\cal O}_8({}^3S_1)\rangle /(2J+1)$
&  0.005 ${\rm GeV}^3$ & 0.014 ${\rm GeV}^3$
\\
\end{tabular}
\end{table}

\begin{table}[b]
\caption{The ratio of the cross section for prompt $\psi + H$ production
to the cross section for prompt $\psi + g$ production with
$H=\psi',\chi_{cJ},\Upsilon(1S),\Upsilon(2S),\chi_{bJ}$, based on direct
$g\to H$ fragmentation contributions only.}
\label{dhcsrat}
\smallskip
\begin{tabular}{cc}
$H$ &
$
\sigma(p \bar{p} \to \psi_{\mu \mu} HX) /
\sigma(p \bar{p} \to \psi_{\mu \mu} gX)
$
\\ \hline
$\psi'_{\mu \mu}$ & $K(g \to \underline{8} {}^3S_1) \,
B(\psi' \to \mu^+ \mu^-)
\langle {\cal O}_8^{\psi'} ({}^3S_1) \rangle = 6 \times 10^{-7} $
\\
$\chi_{cJ}$ & $K(g \to \underline{8} {}^3S_1)  \,
\langle {\cal O}_8^{\chi_{cJ}} ({}^3S_1) \rangle = (2J+1) \;
5 \times 10^{-5} $
\\
$\Upsilon_{\mu\mu}(1S)$ & $K(g \to \underline{8} {}^3S_1) \,
B(\Upsilon(1S) \to \mu^+ \mu^-)
\langle {\cal O}_8^{\Upsilon(1S)} ({}^3S_1) \rangle  = 5 \times
10^{-8}$
\\
$\Upsilon_{\mu\mu}(2S)$ & $K(g \to \underline{8} {}^3S_1) \,
B(\Upsilon(2S) \to \mu^+ \mu^-)
\langle {\cal O}_8^{\Upsilon(2S)} ({}^3S_1) \rangle  = 1 \times
10^{-8}$
\\
$\chi_{bJ}(1S)$ & $K(g \to \underline{8} {}^3S_1) \,
\langle {\cal O}_8^{\chi_{b0}} ({}^3S_1) \rangle  = (2J+1) \; 3 \times
10^{-5} $
\\
\end{tabular}
\end{table}
}

\begin{figure}
\caption{The differential cross section for $p\bar p\to\psi\psi X$
versus $p_T(\psi)$ at the Tevatron.}
\end{figure}

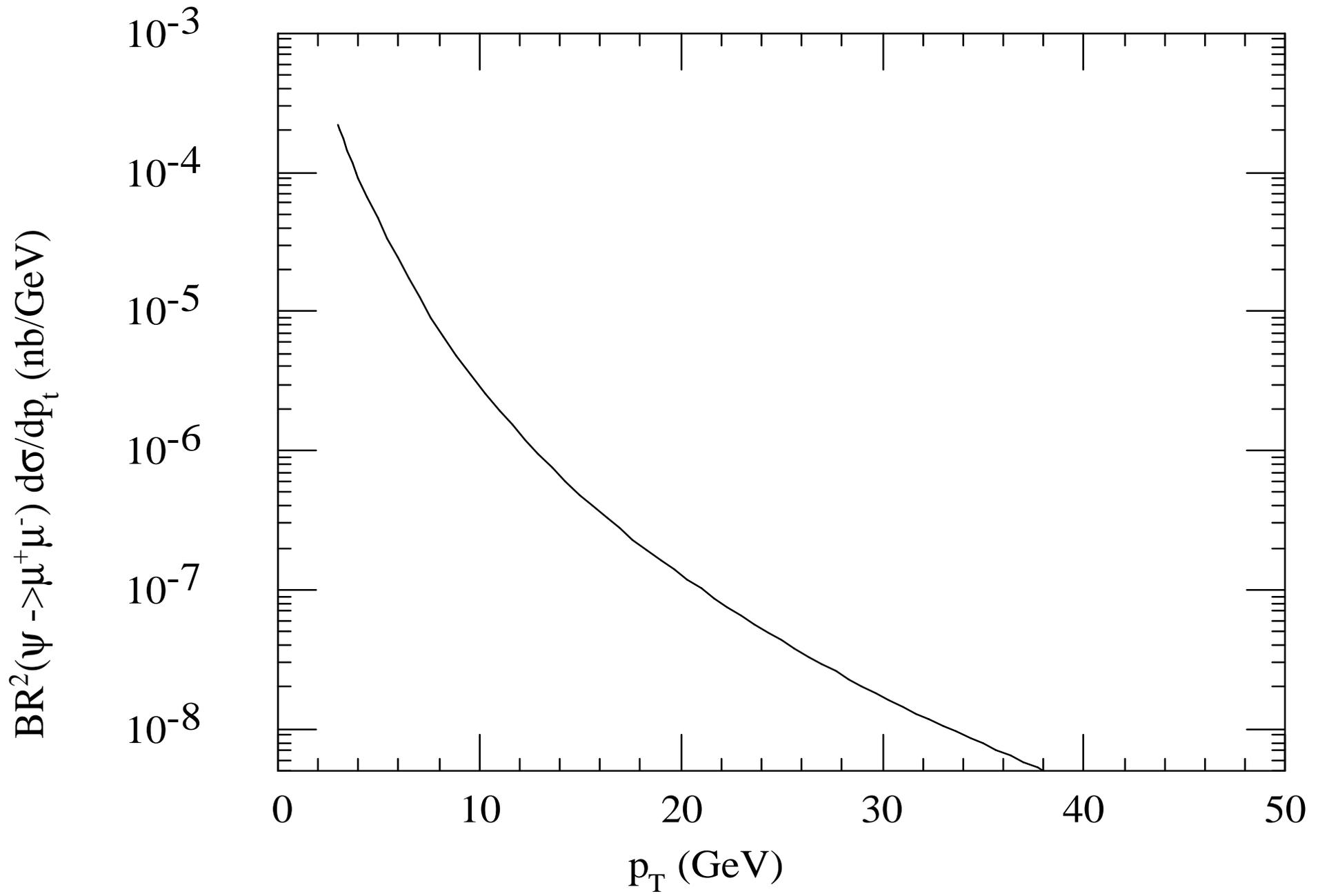
\end{document}